\documentclass{emulateapj}

\usepackage{epsfig}
\usepackage{graphicx}
\usepackage{ifthen}
\usepackage{amssymb}
\usepackage{subfigure}
\def\kms{\,\rm km\,{s}^{-1}}
\def\msun{M_\odot}

\def\beq{\begin{equation}}
\def\eeq{\end{equation}}
\def\pppm{\,\rm P^3M}
\def\omega0{\Omega_{\rm m,0}}
\def\kpc{\,\rm kpc}
\def\mpc{\,\rm Mpc}
\def\LCDM{\Lambda{\rm CDM}}
\def\zs{z_{\rm s}}
\def\zl{z_{\rm l}}
\def\Ds{D_{\rm s}}

\def\lya{Ly $\alpha~$}

\newcommand{\mnras}{MNRAS}
\newcommand{\ba}{\begin{eqnarray}}
\newcommand{\ea}{\end{eqnarray}}
\newcommand{\bi}{\begin{itemize}}
\newcommand{\ei}{\end{itemize}}
\newcommand{\bfi}{\begin{figure}
\epsfxsize=9cm
\epsffile}
\newcommand{\efi}{\end{figure}}
\begin{document}

\title{Detecting First Star Lyman-$\alpha$
  Spheres through Gravitational Telescopes}
\author{GuoLiang Li\altaffilmark{1,2},Pengjie Zhang\altaffilmark{1,2},
 Xuelei Chen \altaffilmark{3}}
\altaffiltext{1}{Shanghai Astronomical Observatory, Nandan Road 80, Shanghai 200030, China; Email: {\tt
(lgl, pjzhang)@shao.ac.cn}}
\altaffiltext{2}{Joint
Institute for Galaxy and Cosmology (JOINGC) of SHAO and USTC}
\altaffiltext{3}{National Astronomical Observatory of China, 20A Datun
Road, Beijing 100012, China}
\shorttitle{}
\shortauthors{Li et al.}

\begin{abstract}
Lyman-$\alpha$ spheres, i.e. regions around the first stars which are  
illuminated by Lyman $\alpha$ (hereafter, \lya) photons and 
show 21cm absorption feature 
against the CMB, are smoking guns at the dawn of the reionization
epoch. Though overwhelming radio foreground makes their
detections extremely difficult, we pointed out that, strong
gravitational lensing can significantly improve their observational
feasibility. Since \lya spheres have $\sim 10^{''}$ 
sizes, comparable to the caustic size of galaxy clusters,
individual images of each strongly lensed \lya sphere often
merge together and form single structures in  
the 21cm sky with irregular shapes. Using high-resolution
$N$-body  $\Lambda$CDM simulations, we found that the lensing
probability to have magnification bigger than $10$ is $\sim
10^{-5}$. This results in $\ga   
10^6$ strongly lensed \lya spheres across the sky, which
should be the primary targets for first detections of \lya
spheres.  Although the required total radio array collecting area for
their detection is large ($\sim 100$ km$^2$), the design of  long
fixed cylindrical reflectors can 
significantly reduce the total cost
of such array  to  the level of the square kilometer array (SKA) and
makes the detection of these very first objects feasible.   

\end{abstract}

\section{INTRODUCTION}
In the concordance cosmology,  first stars and
galaxies formed at redshift $z\sim 30$ in dark 
matter halos of mass $\ga 10^{6}\ M_\sun$. Inefficient gas cooling,
due to the absence of heavy elements, possibly results in first stars
of mass around several hundred solar mass. These first stars emit
numerous near \lya photons directly. 
Furthermore, secondary
\lya photons can be 
generated through the impact excitations by electrons produced by the X-rays
from the high temperature stellar photosphere.
Through frequent \lya scattering in surrounding neutral
hydrogen atoms, the neutral hydrogen
spin temperature  is coupled to its kinetic temperature. Since the
kinetic temperature is lower than the CMB temperature at $z\sim 30$, a
region of 21cm absorption against the CMB develops around each first
stars/galaxies  \citep{Cen06,CM06}. Such regions, 
named ``\lya spheres'' or 21cm absorption ``halos'', are about of
$10^{''}$-$100^{''}$, 
much larger than the first stars/galaxies themselves and  are possibly  the
first distinctive structures in the universe accessible to
observations. As unique features of 
first stars/galaxies,  they will allow 
direct probe of  the very first stage of the reionization
process. This
  approach  is quite complementary to the statistical approach of 
the 21cm observaton, such as measuring 21cm intensity power spectrum and
bispectrum.  It will shed light directly on the reionization
mechanism and could break degeneracies  arising in the modeling of
statistical quantities, e.g.,  degeneracy in various reionization
mechanisms.

However, detections of these first structures are  extremely
difficult, due to overwhelming radio foregrounds and stringent
resolution 
requirement.  For 21cm absorption ``halos'' around stars in the 
$10^8 M_\sun$ dark matter
halos \citep{Cen06}, a total collecting area of $A_{\rm tot}\sim 100$ km$^2$ is
required to resolve them and a $A_{\rm tot}$ of
$\sim 10$ km$^2$ is
required to detect them \citep{Zhang06}. \lya
spheres around the first stars which formed in  $10^6 M_\sun$ dark matter  
halos, as  discussed in \citet{CM06}, are even smaller ($\sim 10^{''}$)
with  weaker signals.  Their detection requires $A_{\rm tot}\ga 10^3$
km$^2$. Such mission would be extremely difficult.

Fortunately, as shown in this paper, strong gravitational
lensing of these \lya spheres makes the first detection
significantly easier. Galaxy clusters, the so called {\it
  gravitational telescopes}, have the power to magnify these
objects by a factor of $10$ or more and thus can significantly improve
the observational feasibility. As estimated in this paper, a 
survey capable of detecting these strongly lensed  \lya
spheres can be built with cost comparable to the square kilometer array
(SKA\footnote{http://www.skatelescope.org/}). 
Such survey will map the 21cm sky to very high accuracy over a large
fraction of the sky and a wide redshift range. It will have
significantly impact on both the fields of  cosmology and reionization. For the
cosmology part, it will  allow precision lensing reconstruction 
through the 21cm background \citep{Cooray04, Pen04,Zahn06} and precision
measurement of the matter power spectrum \citep{Loeb04}.
Furthermore, it will  also resolve 21cm absorption 
'halos', which again have powerful cosmological applications
\citep{Cen06,Zhang06}.

Using strong lensing to detect high $z$ objects is not a new
idea (see \citet{Fort94} for an early review). Using this method,
several high redshift galaxies at $z\sim 6$-$10$ have been detected
\citep{Pello04,Richard06}.   \lya spheres lie at higher
redshifts, so the probability to be highly
magnified is slightly higher.  More importantly, strong lensing of
\lya spheres carry a feature distinctive to that of ordinary
galaxies, explained as follows.

Ordinary galaxies at cosmological distance typically have $\sim
1^{''}$ size,
 this size is usually much smaller than the caustic size of
lens. So images of lensed galaxies are often separated. However,
\lya spheres  have size $\sim 10^{''}$, such a size is comparable 
to the caustic size of galaxies clusters. As a consequence, 
individual images of each strongly lensed
\lya sphere often merge together and form a single structure in 
the 21cm sky with irregular shapes. Although individual images of
larger source usually has lower 
magnification, the merged image can still have a larger
magnification. This effect reduces the
 requirement on $A_{\rm tot}$.  Since there
 are hundreds of billions of 
\lya spheres, even a tiny probability to have a large  
magnification would result in a significant number of strongly lensed
merged images. Interferometer array with 
$A_{\rm tot}\ll 1000$ km$^2$ would be sufficient to
detect these 
 lensed  \lya spheres.

The outline of this paper is as follows. In \S2,
we discuss the  high resolution N-body numerical simulations we use
and the method we 
employ to calculate the lensing probability. In \S 3, we describe the
strong lensing 
 features of \lya
spheres and present the result of  the lensing probability.  In \S 4
we estimate the observational requirements and feasibility and we
summarize our results in \S 5.

\section{NUMERICAL SIMULATION AND LENSING METHOD}

The cosmological model considered here is the concordance 
$\Lambda$CDM model with the dimensionless matter density
$\omega0=0.3$,  the cosmological constant $\Omega_{\Lambda,0}=0.7$, the shape
parameter $\Gamma=\omega0 h=0.21$,  $\sigma_8=0.9$, where $h$ is the Hubble
constant in units of $100\kms\mpc^{-1}$ and we take $h=0.7$.
 A cosmological N-body
simulation with a box size $L=300h^{-1} \mpc$, which was generated
with our vectorized-parallel $\pppm$ code (\citealt{JS02};
\citealt{Jing02}), is used in this paper. The simulation uses
$512^3$ particles, so the particle mass $m_{\rm p}$ is $1.67\times
10^{10}h^{-1}\msun$. The gravitational force is softened with the
$S2$ form (\citealt{HE81}) with the softening parameter $\eta$ taken
to be $30h^{-1}\kpc$. Since strongly magnified images are produced
mostly at radii $\ga 100h^{-1} \kpc$ in the lens plane, the
resolutions are sufficient (see also \citealt{Dalal04}).

Dark matter halos are identified with the friends-of-friends method
using a linking length equal to 0.2 times the mean particle
separation. The halo mass $M$ is defined as the virial mass enclosed
within the virial radius according to the spherical collapse model
(\citealt{KS96}; \citealt{BN98}; \citealt{JS02}).

For a given cluster, we calculate the smoothed surface density maps
using the method of \citet{Li06}. Specifically, for any line of
sight, we obtain the surface density on a $1024 \times 1024$ grid
covering a square of (comoving) side length of $6h^{-1}\mpc$ centered
on each cluster. The projection depth is chosen to be $6h^{-1}\mpc$.
Notice that the size of the region is chosen such that particles
within a few virial radii are included. Particles outside this cube
and large-scale structures do not contribute significantly to the
lensing cross-section  (e.g.  \citealt{Li05}; \citealt{Hen07}). Our
projection and smoothing method uses a smoothed particle
hydrodynamics (SPH) kernel to distribute the particle mass on a 
three-dimension
grid and then integrate along the line of sight to obtain the
surface density (see \citealt{Li06} for detail).
 In this work, the
number of neighbors used in the SPH smoothing kernel is fixed to be
32. We calculate the
surface density along three perpendicular directions for each cluster.
Once a surface density map is obtained, we compute the
cross-section of highly magnified images following the method given in
\citet{Li05}.

 We put the sources at $\zs=30$, a
typical redshift of \lya spheres. The dependence of strong lensing
probability  on $z_s$ at $z_s\ga 10$ is negligible, due to the weak
dependence of source distance on $z_s$ and the fact that no clusters
at redshift beyond several. We simplify the
background sources as spheres with  
diameter $d_{\rm S}=5'',7.5'', 10'', 15'',20''$. This choice of source
size roughly covered the source size in \citet{CM06}. Throughout
  this paper, we have 
neglected the difference in the 21cm brightness temperature across
\lya spheres. Since 
gravitational lensing preserves the 21cm brightness temperature, from 
the mapping between source position and lensing position,
 the temperature distribution can be recovered straightforwardly.

We generate a
large number of background sources within a rectangle box.
 This rectangle is chosen to enclose all the
high-magnification regions($\mu\gtrsim 1.7$ and $\mu< 0$) 
that can potentially form
highly magnified images. Such regions are much larger than
 the size of caustics and provide
a reliable
lensing cross section of image with $\mu \gtrsim 2$. The sources are
located on a regular grid that cover this rectangle with a
resolution of $0.5''$. For each source, ray-tracing is used to find the
resulting image(s). The magnification of each image is defined as
the ratio of the image and source's areas. The resolution in image
plane is kept as $2''$ for sources with $d_{\rm S}=5'',7.5'', 10''$ and 
$4''$ for $d_{\rm S}=15'',20''$, so the total number of pixel is large than 40
for the image with magnification larger than 2. We calculate the
total cross sections of the top 1000$\sim$2000 most massive clusters in each
simulation output. Finally, we obtain the average cross section per unit
comoving volume by: 
\beq \overline\sigma(\zl) = {\sum\sigma_i(\zl)
\over V}, 
\eeq
where $\sigma_i(\zl)$ is the average cross-section of the three
projections of the $i$-th cluster at redshift $\zl$,  and $V$ is the
comoving volume of the simulation box. The optical depth can then be
calculated as: 
\beq \label{eq:tau} \tau = {{1\over{4\pi{\Ds}^2}}
{\int_0}^{\zs}\, dz \, \overline\sigma(z) (1+z)^3\, {dV_{\rm
p}(z)\over dz}}, 
\eeq 
where $\Ds$ is the angular diameter distance
to the source, $\zs$ is the source redshift, and
 $dV_{\rm p}(z)$ is
the proper volume of a spherical shell with redshift from $z$ to
$z+dz$. We used 28 simulation outputs with redshift from 0.2 to 3.
The integration step size is the same as the redshift interval of
simulation output ($d z\approx 0.1$).

\section{Strong lensing probability}
\begin{figure}
{
 \centering
\columnwidth=.99\columnwidth
\includegraphics[width={\columnwidth}]{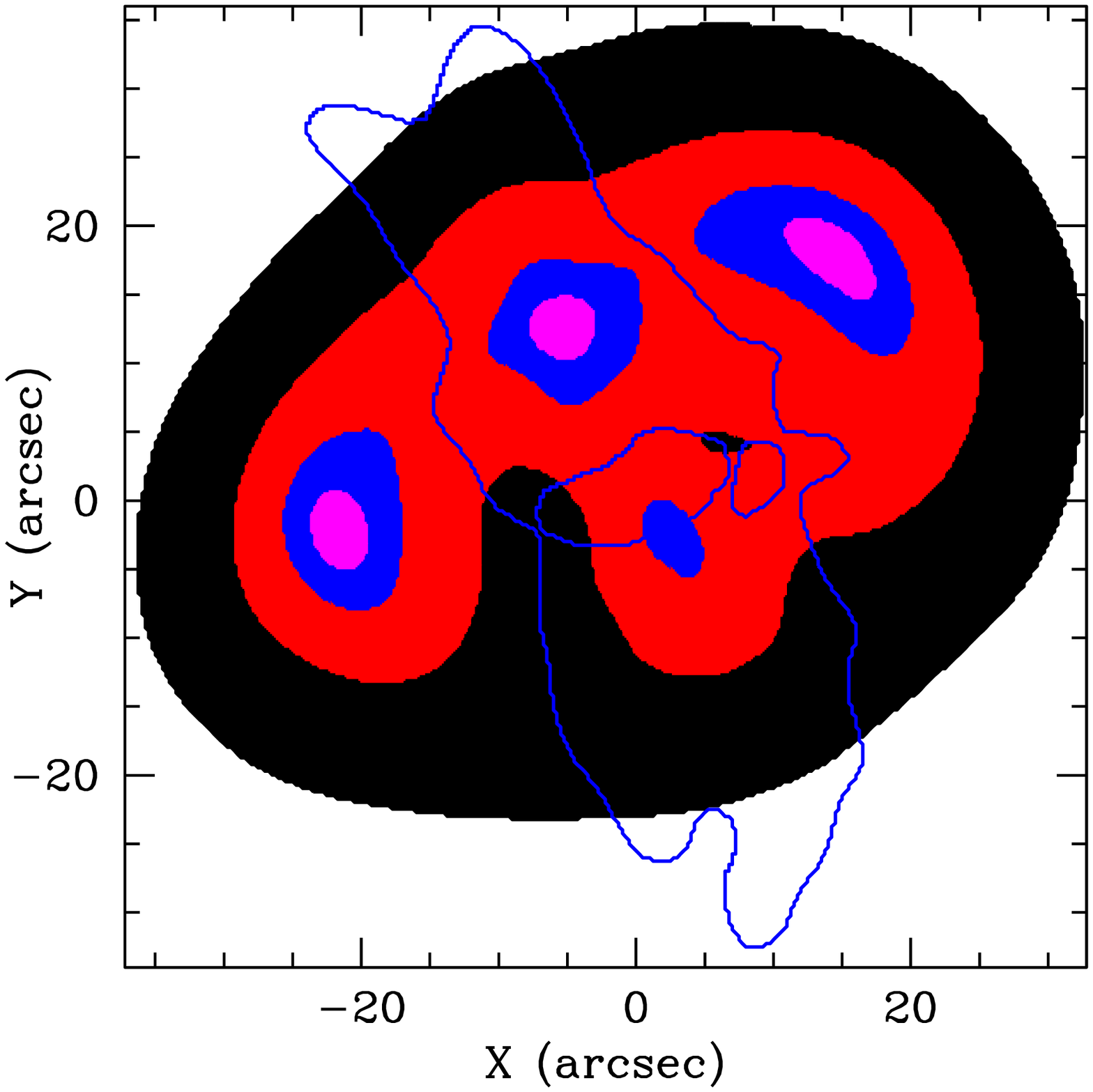}%

\includegraphics[width={\columnwidth}]{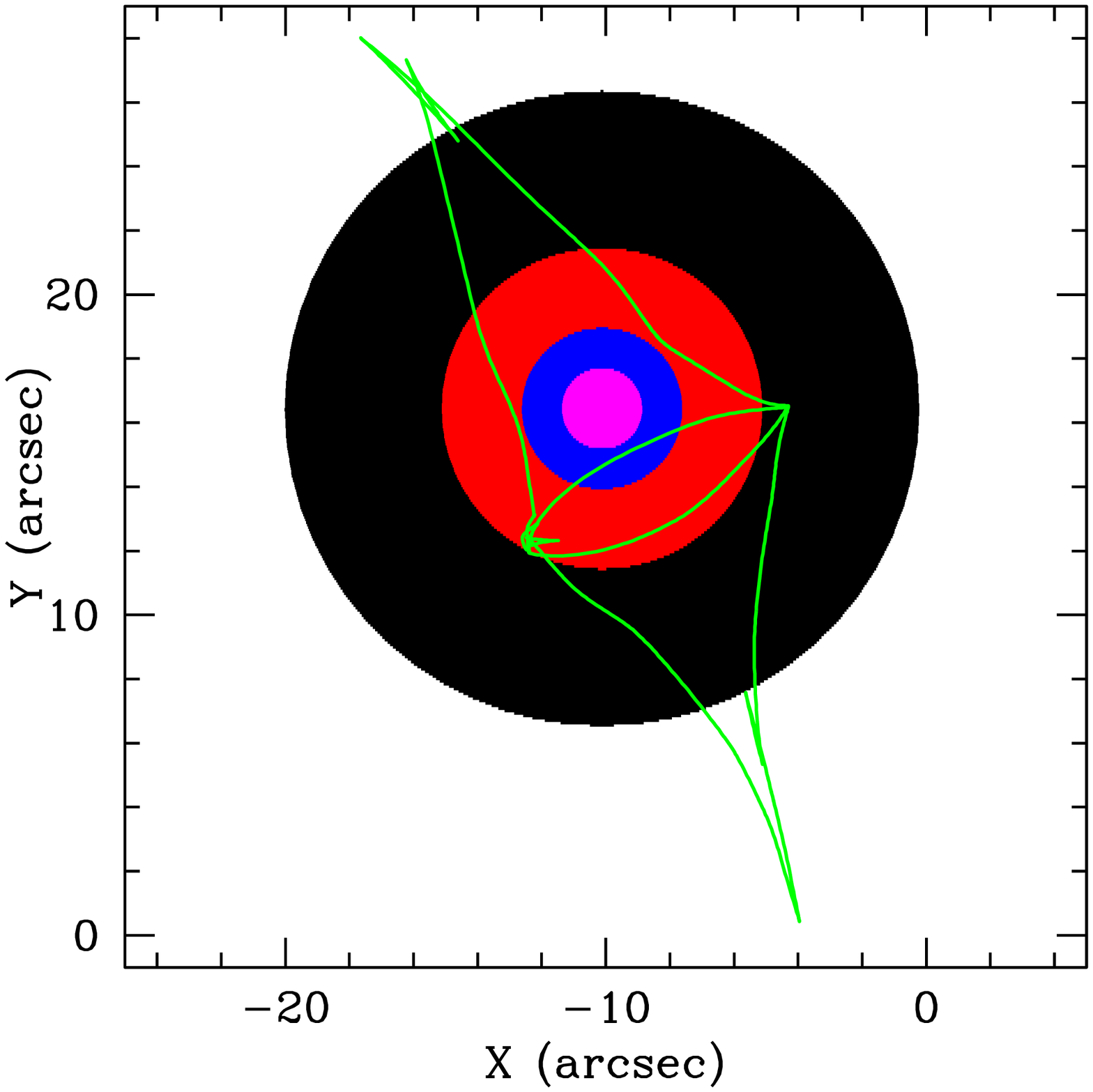}%
}
\caption{The strong lensing signatures of first star \lya 
 spheres. Sources at $z=30$ with different diameter $d_{\rm
 S}=2.5'',5'', 10'',20''$ are shown in purple, blue, red  
and black respectively in the bottom panel. The
 corresponding images are shown in the top panel with same colors. 
The lens is at redshift $z=0.36$ with mass,
 $M_{\rm vir}=1.5\times 10^{15}h^{-1}M_{\odot}$. The blue curves are
 critical curves and the green curves are  caustics. When increasing
 the size of the source to cross the caustics, 
 some separate lensed images will merge together. For a sufficiently
 large source, all lensed images will merge together and form a single
 structure with irregular shape. This feature is distinctive to
 ordinary strong lensing of galaxies. The resulting single lensed
 images will most likely be the first detected \lya
 spheres. } 
\label{fig:1}
\end{figure}

To illustrate the signatures of lensed \lya spheres, we
choose a cluster at $z=0.36$ as the lens. By varying the size of the
\lya sphere, we find an interesting lensing feature. Usually, the 
magnification decreases quickly with expanding source size, because 
the magnification which close to the caustic 
follows $\sim \mu\propto y^{-1}$, where $y$ is the distance
to caustic from 
inner. On the other hand, there are also two effects to increase the high 
magnification region. When a source is close to but outside of the caustic, the 
magnification is modest for a point source, but for an extended source
which is large enough to cover the caustic, a highly magnified
image results. So the finite source can extend the high magnification region 
to the outside of the caustics.  
When a source is inside but faraway from the caustics, it will create 
multiple images 
with also modest magnifications, but if the source is large enough to
match the  caustic(s), some or all of these images will merge
together, then the 
final magnification is still very high. Fig. \ref{fig:1} shows such
merger effect 
of images. The lens is at redshift 0.36 with mass, $M_{\rm vir}=1.5\times 
10^{15}h^{-1}M_{\odot}$ and the sources are fixed at redshift 30. Notice,
there are no overlaps for a merged image. The incomplete images 
connect together along the critical curve(s) and build up such a larger image.

 Since the typical size of \lya sphere is $\sim10^{''}$ and is
indeed comparable to the size of the caustics, it is frequent that
lensed \lya sphere will show up as a single merged
image. The area of this image is a factor of $\mu$ of the source,
where $\mu$ is the total magnification.

\begin{figure}
{
 \centering
\columnwidth=.95\columnwidth
\includegraphics[width={\columnwidth}]{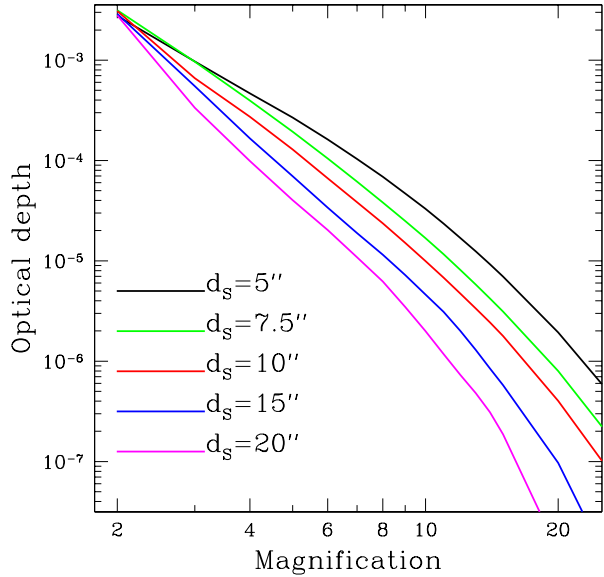}%
}
\caption{The optical depth as a function
of magnification. Different color of line corresponding different size of 
source respectively. The optical depth for $\mu\ge2$ virtually does not depend on the source
size. This is because the region with $\mu\ge 2$ in source plane is 
much larger than the size of caustic or the size of source.
 Hence, the cross-section with $\mu\ge2$ is dominated by the low magnification
 region and the optical depths are almost same for the 
point sources and the finite sources.
 } 
\label{fig:2}
\end{figure}

Fig. \ref{fig:2} shows the optical depth as function of 
magnification. For typical size of \lya
spheres $\sim 10^{''}$, there is $10^{-5}$ probability to produce
$\mu>10$ merged images. Since there are hundreds of billions
\lya spheres, we expect to find hundreds of thousands lensed
\lya spheres with $\mu>10$.

\subsection{Model uncertainties}
\begin{figure}
{
 \centering
\columnwidth=1.\columnwidth
\includegraphics[width={\columnwidth}]{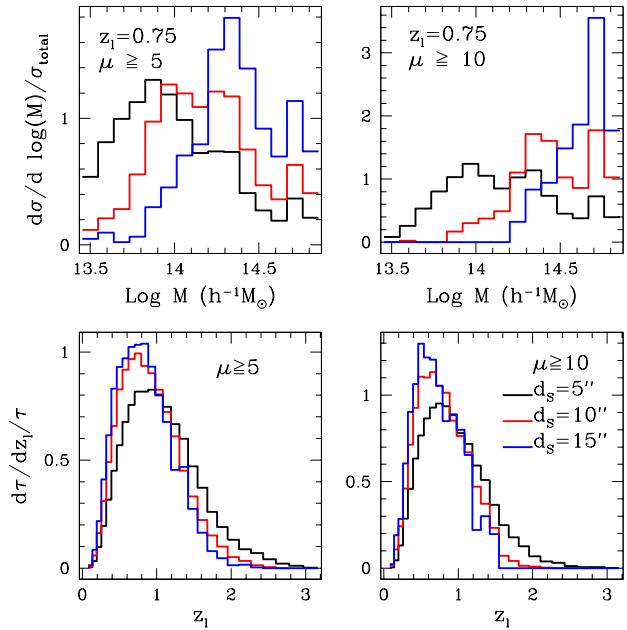}%
}
\caption{The two top panels show the differential cross 
section as a function of logarithm lens mass. The magnification threshold is
 labeled in each panel and the representing of color for each histogram is
 as same as in top and bottom panels. The total area under each curve
is normalized to unity. The two bottom panels show the differential probability 
distribution for the optical depth as a 
function of lens redshift. The optimal lens redshift changes from 0.5 to 1 
following the changing of source size and magnification threshold. Larger source size and
magnification threshold causes lower optimal lensing redshift.} 
\label{fig:3}
\end{figure}

We further investigate the completeness of our simulated cluster
sample in mass and redshift range and thus the robustness of our
predicted $P(\mu)$.  Lower panels of fig. \ref{fig:3}
show the differential probability distribution for the optical depth  
as a function of lens redshift, for various source sizes and
magnification thresholds. The optimal lens redshift for larger source
size and larger  magnification 
threshold is lower. For the source size $5^{''},10^{''}, 15^{''}$ and
$\mu\geq 5,10$ investigated, the optimal lens
redshift  varies between 0.4 and  1. Thus cluster at redshift around
0.4 to 1 will be most likely responsible for strong lensing of
\lya spheres.

Top panels of  fig. \ref{fig:3} show differential cross 
section as a function of logarithm lens mass, 
$d\sigma/d\rm log(M)/\sigma_{\rm total}$,  where $M$ is the virial
mass of individual lens cluster and $\sigma_{\rm total}$ is the
summation of the mean cross-sections of all clusters.  Since the
optimal redshift is between $0.4$ and $1.0$, 
we choose a median redshift $z=0.75$. For this redshift, we select 
the top 2000 most massive clusters in the simulation box. 
 The low mass cut is about $3\times 10^{13}$ and the corresponding number
of particle within in virial radius is $\sim 2000$. 
The small cluster has too small caustic(s) to highly magnify a big source.
But when the source size is smaller, the source can have non-negligible
probability to be  strongly lensed by
a less massive cluster, which may not be well resolved in our
simulation. Meanwhile, when estimating the lens optical depth
of small $\mu$, contribution from less massive clusters may be also
non-negligible for their huge abundance. Thus, for small $\mu$ and 
small source size, our
cluster sample can be incomplete. Top left panel of Fig. \ref{fig:3}
shows that this is the indeed case for source size $5^{''}$ and
$\mu\geq 5$. However, for the more interesting case, $10^{''}$ and/or
$\mu \geq 10$, contribution from clusters less massive than
$10^{13.5}h^{-1}\msun$  is negligible. Thus, our cluster sample is
complete at low-mass end for $10^{''}$ sources. The last figure tells us that
 the optical depths are complete at low-mass end for $\mu\ge 10$ 
for any size of source. Particularly, the lensing probability
 is $10^{-5}$ at this magnification threshold.

 On the other hand, Fig. \ref{fig:3} implies that clusters more massive than
$10^{15}h^{-1}\msun$ still have non-negligible contribution. Due to
the finite box size, massive clusters are not fairly represented in our
simulation. The lensing probability of $\mu>10$ may be
under-estimated by a factor no larger than $2$ for the lack of very massive
 clusters in our simulation.

 As seen from Fig.~\ref{fig:3}, halos less massive than
$10^{13.5}h^{-1}\ \msun$ have negligible contribution to $\mu>10$
lensing. Thus, we do not expect galaxies, which are not included in
our simulation, can affect the above result significantly. This can be
further understood as follows. For an
isothermal distribution of matter in galaxies, the Einstein radius
$\theta_E=(\sigma_v/186 
{\rm km}\ s^{-1})^2D_{\rm LS}/D_S$ arcsec. Since it is much smaller
than the size of \lya spheres, it is not surprising that
they do not contribute much to strong lensing of \lya
spheres. 

We have to say that the lensing probability we give here is just a low limit 
in the concordance $\LCDM$ model. The baryon effect is not taken into 
account. In the real universe,
baryons account for roughly 20\% of the total mass, which
can cool (form stars) and sink toward the centers of clusters.
The radiative cooling likely has two effects: it will
increase the concentration of baryons at the center of clusters,
and at the same time, make the clusters more spherical (e.g.
\citealt{Dubinski94, Kaz04}). The former increases while the latter 
decreases the lensing cross-sections, and so the overall influence
depends on which effect dominates. \citet{Mene03} and \citet{Puchwein05} 
found that the increase in the lensing cross-section of giant arc 
 due to baryon is quit modest, by a factor of $\la 2$ for massive cluster
and the the increase becomes larger for low-mass cluster. So, we can 
estimate the probability as  $\sim2\times 10^{-5}$ if we take into account 
the baryon.

From the two top panels in fig. \ref{fig:3}, we can see our cluster sample
is usually incomplete at low-mass end (for small source and low 
magnification threshold) or at high-mass end (for large source and high 
magnification threshold). This leads to an under-estimation of the lensing 
probability by a factor of $\la50$\%. In this paper,
we use thin lens approximation but the large-scale structures
 along the line of sight also should have additional contribution to the
lensing probability although this is very modest 
(e.g., \citealt{Li05}; \citealt{Hen07})

On the other hand, recent results from the three-year data of 
 the Wilkinson Microwave Anisotropy Probe (WMAP)
 prefers lower $\sigma_8=0.74$ and $\omega0=0.238$. 
than that in the concordance $\LCDM$ model
 \citep{Spergel06}. Since the abundance of 
 massive clusters decreases dramatically in this low $\sigma_8$
 universe,  the number of giant arcs is a factor of $\sim 6$ lower
 than that in the concordance model \citep{Li06b}. 
 Roughly, we expect a similar decrease for the lensing 
probability of the \lya spheres. Despite of the baryon and 
other contributions we discussed above, we estimate the optical depth as
$~10^{-5}/6\approx 2\times 10^{-6}$. 

The total number of \lya
spheres across the sky is $\sim 10^{11} - 10^{13}$, depending 
on the cosmological parameters adopted and the redshift range
\citep{CM06}.  Since these  \lya
 spheres have typical diameter $10^{''}$, some of them apparently overlap along
 the line of sight. However, when they are visible 
 in 21cm, they are also well separated in redshifts. 
The density of these spheres is between
 0.1 to 1 comoving $\mpc^{-3}$ at redshifts 30, c.f. Fig. 5 of 
\cite{CM06}, while their diameter is about 20 kpc physical, or 600
kpc comoving, so they are well seperated and can be distinguished at
different frequencies. As will be shown in next section, the optimal
bandwidth of observation is about 20 kHz, which could easily
distinguish these different spheres along the line of sight. 
There is some chance that 
within the \lya sphere another first star form, and produce 
another \lya sphere which overlaps with the first, 
the two spheres would be physically
merged. This is however not common. The probablity of finding another 
halo within a \lya sphere is about 0.3 (Fig. 8, op.cit.), 
but given the short life time of the first stars, the chance of the
two stars shine at the same time is even smaller.
Of course, eventually as
more and more of these spheres appear, 
a homogeneous \lya background is built up. At this point,
the \lya flux induced spin temperature difference
diminishes, and the \lya sphere become invisible. This
happens before the physical overlap of the spheres, at a filling
factor of about 0.1(see the discussion on the build up of
\lya background in \S 4.1 of \cite{CM06})

Given the probability $10^{-5}$ to be
amplified by a factor of $10$ or more, there will be at least 
$10^6 - 10^8$ of \lya spheres which are strongly lensed 
in the concordance model. A radio array with  total
collecting area $A_{\rm tot}\la 100$ km$^2$ while keeping cost
comparable to SKA is
promising to detect many of them in years survey as the discussion bellow.

\section{Observational feasibility}
A large $\mu$ will significantly reduce the requirement on
the resolution and r.m.s. noise of a radio survey. The exact number
depends on the detailed   array configuration. Without loss of
generality, we focus on a configuration of  
$N\times N$ arrays homogeneously distributed over a square with side length 
$L$. For an optimal detection, the resolution  $\theta_p\simeq
\lambda/L$ should be similar to the size of the 
\lya sphere, Here,  $\lambda=21(1+z)$cm  
is the redshifted wavelength of the 21cm line.  The 
residual noise  per resolution pixel (with area $\theta_p^2$) is
\ba
\label{eqn:sigmaT}
\sigma_T&\simeq&
\frac{T_{\rm sys}}{\sqrt{\Delta \nu
    t}}\frac{1}{f_{\rm cover}}\\
&\simeq& 20 {\rm mk} \frac{T_{\rm
    sys}}{3000 {\rm K}} \left(\frac{\Delta \nu}{10{\rm
    kHz}}\frac{t}{\rm year}\right)^{-1/2} \frac{26\%}{f_{\rm cover}}
\nonumber
\ea
Here, $f_{\rm cover}=A_{\rm tot}/L^2$ and
$A_{\rm tot}$ is the total collecting area. The bandwidth $\Delta
\nu$ is related  
to the comoving separation $r$ by $\Delta \nu\simeq 42 {\rm kHz}
 [31/(1+z)]^{1/2} [d/h^{-1}{\rm Mpc}]$. For the typical signal $\sim
 -100 {\rm mk}$ and typical comoving \lya sphere diameter $d\sim 0.6
 $ Mpc (physical diameter $\sim 20$ kpc), assuming the integration $t$
 = 1year, $A_{\rm 
   tot}\sim 1000$ km$^2$ is required to detect unlensed \lya
 spheres \citep{CM06}.  

However, a factor of $\mu$ amplification in the area relaxes the requirement on
resolution and thus the requirement on
$L$ by a factor of $\mu^{1/2}$. Since the differential brightness
temperature does  not change, to obtain the same S/N, 
$A_{\rm tot}$ can be reduced by a factor of
$\mu$, if $\sigma_T$ and $f_{\rm cover}$ are kept fixed. Thus, for
those highly amplified \lya spheres with  $\mu>10$, a radio
array with  $A_{\rm tot}\la 100$ km$^2$ is sufficient.

A radio array with $A_{\rm tot}\sim  100$ km$^2$ can be built of fixed
parabolic cylindrical reflectors \citep{Peterson06} with reasonable
cost.  Such  design can 
reduce the cost to \footnote{Private communication 
with Ue-Li Pen.} $\sim 10\$/$m$^2$.  
 Given this number, the total cost can be controlled to be  
around one billion dollar, comparable to the estimated total cost of
SKA, which has unit cost of 
$\sim 1000\$/$m$^2$. This design has large field of view
(several thousand  square degree at $z=30$) and can scan
at least half the sky through the Earth rotation. Since the
field view is large, the telescope does not need to point to specific
regions of known galaxy clusters, although these regions are of
primary interest in the process of data analysis. With these advantages, this
design  is appropriate
for the detection of \lya spheres in blank sky.

\section{Discussions and summary}

The cluster which serves as gravitational lens may have radio
emission, this could complicate the situation. However, we expect that
the main mechanisms of radiation in the clusters--synchrotron and
free-free-- produce smooth frequency spectrum, so we should be able to
disentangle these from the lensed 21cm signal. The cluster radio
emission does increase the sky background noise, hence the boost in
the signal to noise ratio is less than expected from the lensing
magnification. However, radio observation of nearby clusters show that
not all clusters have radio strong emission, if we avoid those with 
FR-II radio sources, then many clusters have radio emission power
less than $10^{25} W/s$ at 20 cm \citep{DY02}. 
For a cluster at cosmological distance ($10^3$ Mpc), the corresponding 
brightness temperature of such a cluster is much less than 
the galactic synchrotron foreground, so the increase of noise would
not be significant.

  Ideally, one wants to de-lens individual
  \lya spheres to directly measure their intrinsic
  properties.  However, clusters with $M_{vir} \sim 10^{14}h^{-1}M_{\odot}$ at 
$z\sim1$ (Fig. 3) contribute a non-negligible fraction of  
  the important lenses. These cluster  are generally sub-critical for
  sources at $z\sim 1-3$, faint in X-rays and SZ, and their mass
  reconstruction would therefore be difficult.  Furthermore, the angular
  resolution we consider  here is comparable to the size of the (lensed)
  spheres. This makes delensing very difficult, even if 
  detailed lens information is given. 

  In the worst case that delensing of individual \lya
  spheres is not feasible, their direct detection
   is still valuable. The direct observable, the number of detected
  (strongly lensed) \lya spheres as a function of 
  redshift, is already valuable for understanding the history of first
  star formation and reionization
  mechanism. By the time of successful 
  \lya sphere detection, galaxy and 
  cluster surveys will be significantly advanced. Cluster abundance and lensing
  efficiency are promised to be well measured through
  observations of strong and weak lensing of 
  galaxies, as well as the X-ray and SZ effect of clusters
  Given these lens information, the prediction of  the
  detected number of spheres  then relies only on the 
  source property. Comparing the prediction with observations
  will then reveal the
  nature of the reionization process.

As a summary, galaxy cluster is a powerfully telescope to magnify first
star \lya spheres. Since the size of \lya spheres
is comparable to the caustic size of clusters, the lensed images often
merger together. This distinctive lensing feature  significantly improves
the observational feasibility of these objects. 
Clusters with $M_{\rm
  vir}\sim 10^{14.5}h^{-1} \msun$ in the
redshift range $[0.4,1]$ are responsible for most of the strong
lensing events (fig. \ref{fig:3}). Regions where these clusters cover
should be the primary locations looking for \lya spheres in
the reconstructed 21cm brightness temperature maps. Cylindrical
reflector array with 
$A_{\rm tot}\sim 100$ km$^2$ and reasonable cost comparable to that of
SKA should be able to detect at least $10^5$-$10^8$ strongly lensed 
\lya
spheres and open a window into the very first objects in the
Universe.

\section*{Acknowledgment}
We thank Ue-Li Pen for helpful information on the design of cylindrical
 reflectors. We thank Yipeng Jing, Shude Mao, XiangPing Wu and the anonymous
 referee for useful
 comments. This work is supported by grants from the
NSFC (No. 10373012, 10533030, 10525314, 10533010), the 
Shanghai Key Projects in Basic research (No. 04JC14079 and 05XD14019),
 and the CAS grant KJCX3-SYW-N2.

\end{document}